\documentclass[pdflatex,sn-mathphys-num]{sn-jnl}% Math and Physical Sciences Numbered Reference Style

\usepackage{graphicx}%
\usepackage{multirow}%
\usepackage{amsmath,amssymb,amsfonts}%
\usepackage{amsthm}%
\usepackage{mathrsfs}%
\usepackage[title]{appendix}%
\usepackage{xcolor}%
\usepackage{textcomp}%
\usepackage{manyfoot}%
\usepackage{booktabs}%
\usepackage{algorithm}%
\usepackage{algorithmicx}%
\usepackage{algpseudocode}%
\usepackage{listings}%
\usepackage{cleveref}
\usepackage{todonotes}
\usepackage{graphicx}
\usepackage{tcolorbox}
\definecolor{hellblue}{RGB}{220, 230, 241}%{173, 216, 230}
\tcbset{
    colback=hellblue,
    colframe=hellblue,
    arc=2mm,
    boxrule=0pt,
}

\usepackage{mdframed}

\theoremstyle{thmstyleone}%
\newtheorem{theorem}{Theorem}%  meant for continuous numbers
\theoremstyle{thmstyletwo}%

\theoremstyle{thmstylethree}%

\makeatletter
\def\url#1{} % Suppresses the URL
\def\doi#1{}  % Suppresses the DOI
\makeatother

\begin{document}

\title[]{Statistical Guarantees for Graph Neural Networks}
\title[]{Statistical Analysis of Graph Neural Networks}
\title[]{Statistical Frameworks to analyse Graph Neural Networks}
\title[]{Theoretical Frameworks to analyse Statistical Generalisation of Graph Neural Networks}
\title[]{Different Statistical Perspectives for Understanding Generalisation in Graph Neural Networks}
%\title[]{Different Perspectives towards Understanding Statistical Error of Graph Neural Networks}


\author[1]{\fnm{Nil} \sur{Ayday}}\email{nil.ayday@tum.de}

\author[2]{\fnm{Mahalakshmi} \sur{Sabanayagam}}\email{mahalakshmi.sabanayagam@adelaide.edu.au}

\author*[1]{\fnm{Debarghya} \sur{Ghoshdastidar}}\email{ghoshdas@cit.tum.de}

\affil[1]{\orgname{Technical University of Munich}, \orgdiv{School of Computation, Information and Technology},  \orgaddress{\street{Boltzmannstr. 3}, \city{Garching}, \postcode{85748}, %\state{Bavaria}, 
\country{Germany}}%
}

%\affil[2]{\orgname{Konrad Zuse School of Excellence in Reliable AI}, \orgaddress{\street{Walther-von-Dyck-Str. 10}, \city{Garching}, \postcode{85748}, %\state{Bavaria}, 
%\country{Germany}}}

\affil[2]{\orgname{Australian Institute for Machine Learning, Adelaide University}, %\orgdiv{College of Engineering and Information Technology}, \orgaddress{\street{Adelaide City Campus East}, 
\city{Adelaide}, \postcode{SA 5000}, %\state{State}, 
\country{Australia}}%
%}


\abstract{Graph Neural Networks (GNN) are currently the most popular approach for learning and prediction on graph-structured data and are deployed in various fields, from social network analysis to drug discovery. However, there is limited mathematical understanding of the performance of GNNs. We discuss the various perspectives used to study statistical generalisation in GNNs. We identify three broad frameworks. The first approach, rooted in learning theory, relies on uniform convergence bounds and the complexity of the hypothesis class of specific GNN architectures. This approach also builds on the expressivity of GNNs, typically studied through the lens of graph isomorphism tests. The second principle is to simplify the neural architecture by analysing GNNs under the asymptotics of infinitely many parameters or infinite graph size. This approach approximates GNNs using Gaussian processes, neural tangent kernels or graphon neural network operators, which allow studying the generalisation or stability of trained GNNs. The third framework studies GNNs under random graph models, often the contextual stochastic block model, and derives non-asymptotic error rates using tools from high-dimensional statistics. We highlight some key theoretical results and discuss a few limitations and open research questions for each perspective.}




\maketitle

\section{Introduction}\label{sec1}

%The Introduction section, of referenced text \cite{bib1} expands on the background of the work (some overlap with the Abstract is acceptable). The introduction should not include subheadings.

%Springer Nature does not impose a strict layout as standard however authors are advised to check the individual requirements for the journal they are planning to submit to as there may be journal-level preferences. When preparing your text please also be aware that some stylistic choices are not supported in full text XML (publication version), including coloured font. These will not be replicated in the typeset article if it is accepted. 

The surge of interest in Graph Neural Networks (GNNs) stems from a fundamental shift in how we approach computationally intractable graph problems. For decades, search or learning problems on graphs, such as \emph{community detection}, \emph{the travelling salesperson problem}, \emph{protein folding}, etc., were primarily viewed as combinatorial optimisation problems. 
The trend in recent years has been to reframe these challenges as learning problems, typically solved using neural networks. For example, \citeauthor{DBLP:conf/iclr/ChenLB19}~ \cite{DBLP:conf/iclr/ChenLB19} ``learn'' the community detection problem by training a GNN on random graphs with communities generated from a stochastic block model \cite{DBLP:journals/jmlr/Abbe17}. 
Notably, GNNs allow more flexibility than standard neural network architectures, such as feedforward, convolutional, and recurrent networks. While standard architectures operate on fixed grid-structured data, such as 2D or 3D-arrays for images, GNNs are capable of operating on arbitrary non-Euclidean topologies, making them suitable for applications like molecular structure prediction. GNNs also allow for the design of predictors that are permutation-equivariant or permutation-invariant.  Hence, GNNs are synonymous with the field of geometric deep learning \cite{bronstein2021geometric} and are deployed in various fields \cite{DBLP:journals/tnn/WuPCLZY21,DBLP:journals/corr/abs-2502-16533}.

The theoretical analysis of GNNs is in a nascent stage, and diverse perspectives have evolved regarding the understanding of the performance of various GNN models. The different theoretical views are natural since one can study GNNs through the lens of graph theory, random graphs, non-Euclidean geometry, statistical learning theory, deep learning theory, etc. We focus on the statistical analysis of GNNs, which asks: \emph{What is the test error (or generalisation error) of a GNN trained with samples from a statistical law?}
Three broad frameworks are used to study generalisation in GNNs:
\begin{itemize}
    \item[1.] \emph{Statistical learning theoretic framework:} 
    Statistical learning theory provides the classical tools to study generalisation for any machine learning model and has been used in early statistical guarantees for GNNs. Such guarantees are useful if one can show that the GNN models can learn complex functions. Hence, this framework is related to the research on the expressive power of GNNs, which we also discuss. 
    \item[2.] \emph{Extension of deep learning theory to GNNs:}
    Classical learning theory often fails to explain the performance of trained neural networks (NN), for example, \emph{benign overfitting} or \emph{neural scaling laws}.
    This has led to new tools for the asymptotic analysis of NNs, deriving nonparametric or operator limits of NNs. Recent work has extended such analysis to study GNNs, providing deeper insights.
    \item[3.] \emph{Statistical network model viewpoint:} 
    The above frameworks are not adapted to the statistical nature of the graph data. A recent direction studies GNNs under specific random graph models, such as the contextual stochastic block model, using tools from high-dimensional statistics. Despite focusing on specific graph models, these results provide precise generalisation error rates and explain empirical observations.
\end{itemize}
Before delving into the details of the above statistical frameworks, we formally describe GNNs, associated learning problems, and the notion of generalisation.

\begin{figure}[t]
\centering
\includegraphics[width=0.9\linewidth]{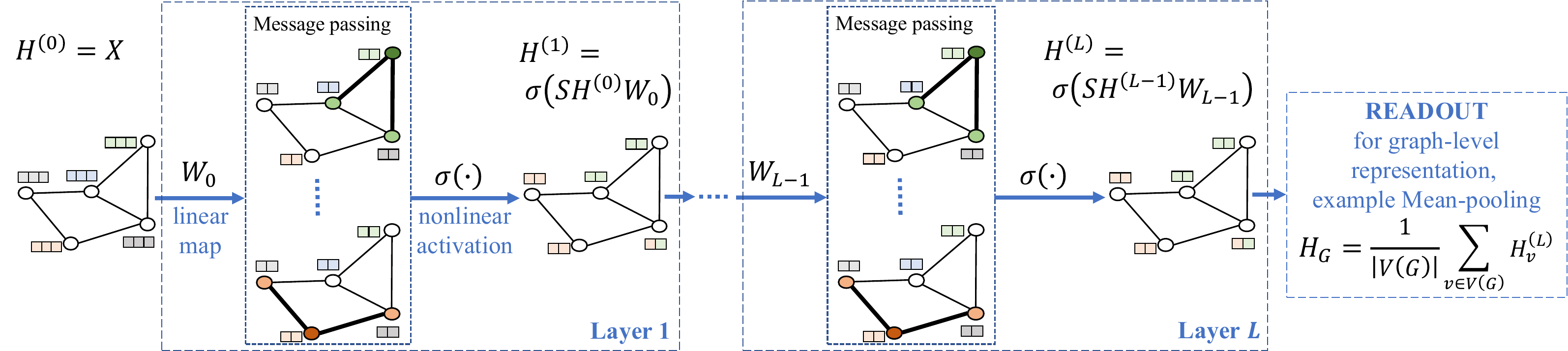}%
%\newline%
%\framebox{\includegraphics[width=0.98\linewidth]{sn-article-template/figures/GNN-WL-example.pdf}}%
\caption{$L$-layer GNN architecture, with expressions specific to graph convolutions (GCN) \cite{DBLP:conf/iclr/KipfW17}. READOUT layer, used for graph-level tasks, combine node representations to a graph representation.
%\textbf{(bottom row)} Illustration of how the GNN layers (left) operate on the 5-cycle graph, $C_5$, and the path graph with 5 nodes, $P_5$. For the GNN, we assume that node features are 1 for every node. 
%The illustration on rights compares the GNN updates with iterations of the Weisfeiler-Lehman algorithm (1-WL) for checking graph isomorphism (see Section \ref{sec:gen_bounds}). Further GNN specifications and the Hash functions in 1-WL are noted. 
}
\label{fig:GNN}
\end{figure}

\paragraph{Graph neural network (GNN)}
The GNN literature typically considers graphs with additional node features. Hence, a graph on $n$ nodes is usually represented by $G=(A,X)$, where $X \in \mathbb{R}^{n\times d_0}$  with the $i$-th row representing the $d_0$-dimensional node features of the $i$-th node, and $A \in [0,1]^{n\times n}$ is the (weighted) adjacency matrix. Let $D$ denote the diagonal matrix  with $D_{ii}=\sum_{j=1}^n A_{ij}$. We discuss only static, undirected graphs.
A $L$-layer GNN iteratively computes the node representations. The $\ell$-th layer computes $H^{(\ell)} \in \mathbb{R}^{n \times d_{\ell}}$, where row $H^{(\ell)}_i \in \mathbb{R}^{d_\ell}$ is a representation of node-$i$.  
Starting with $H^{(0)} = X$, a GNN computes
\begin{equation}
H^{(\ell)}_i = f\left(H^{(\ell-1)}_i,\left\{H^{(\ell-1)}_j, A_{ij} ~\forall j:A_{ij}>0\right\}\right), \qquad \text{for } \ell=1,\ldots, L,
\label{eqn:MP-GNN}
\end{equation}
where $f(\cdot)$ could be a general non-linear function that aggregates the representations in the local neighbourhood of node-$i$, obtained from the previous layer $H^{(\ell-1)}$, along with the edge information $A$. Since the aggregation $f(\cdot)$ essentially acts like ``message-passing'', GNN architectures with the update-rule \eqref{eqn:MP-GNN} are broadly known as message-passing GNNs (MP-GNN). One of the most popular examples of MP-GNN is a graph convolutional network (GCN) \cite{DBLP:conf/iclr/KipfW17} (see Figure \ref{fig:GNN}) with the update rule simplifying to 
\begin{equation}
H^{(\ell+1)} = \sigma\left(S H^{(\ell)}W_{\ell}\right), \qquad \text{for } \ell=0,1,\ldots, L-1,
\label{eqn:GCN}
\end{equation}
where $W_\ell\in\mathbb{R}^{d_{\ell}\times d_{\ell+1}}$ is trainable, $\sigma(\cdot)$ is an entry-wise non-linear map, like ReLU$(z) = \max\{z,0\}$, and the convolution operator $S\in\mathbb{R}^{n\times n}$ is a function of the adjacency $A$, such as symmetric normalised adjacency $S_{sym}= (D+I)^{-1/2}(A+I)(D+I)^{-1/2}$ or row-normalised adjacency $S_{row} = (D+I)^{-1}(A+I)$. The addition of $n\times n$ identity matrix $I$ ensures that each node retains its information. 
The updates \eqref{eqn:MP-GNN}--\eqref{eqn:GCN} provide nonlinear node representations $H^{(L)}$, which can be directly used for prediction problems on the nodes. To obtain a single representation of the graph $G$, a permutation-invariant read-out function is used: $H_G = \text{READOUT}\left(\left\{H^{(L)}_i:i=1,\ldots,n\right\}\right)$, such as the mean, the sum, or the maximum of $H^{(L)}_i$ across all nodes.
Several alternatives exist. Graph attention network (GAT) \cite{DBLP:conf/iclr/VelickovicCCRLB18} replaces $S$ in \eqref{eqn:GCN} with a learnable convolution in each layer
\begin{equation}
S^{(\ell)}_{ij} = \frac{\exp\left(\tilde{\sigma}\left(\left[H_i^{(\ell)}W_{\ell} ~~H_j^{(\ell)}W_{\ell}\right]  v_{\ell}\right)\right)}{\sum_{k\in\{i\}\cup\{k':A_{ik'}>0\}}\exp\left(\tilde{\sigma}\left(\left[H_i^{(\ell)}W_{\ell} ~~H_k^{(\ell)}W_{\ell}\right]  v_{\ell}\right)\right)},
\label{eqn:GAT}
\end{equation}
parameterized by trainable $v_\ell\in\mathbb{R}^{2d_{\ell+1}}$, and non-linear activation $\tilde\sigma(\cdot)$. Thus, the message-passing in \eqref{eqn:GAT} is a softmax ``attention'' within a neighbourhood, providing improved performance in practice.
%
%: the graph attention network (GAT) \cite{DBLP:conf/iclr/VelickovicCCRLB18} is practically useful as it learns the convolution operator $S$, while the graph isomorphism network (GIN) \cite{DBLP:conf/iclr/XuHLJ19} has theoretical significance due to its connection to graph isomorphism tests. 
There exist other architectures beyond MP-GNNs, such as spectral GNNs, where the update rule is in terms of the graph spectra, whereas graph transformers learn global interactions. 
We refer to \cite{bronstein2021geometric,DBLP:journals/tnn/WuPCLZY21,DBLP:journals/corr/abs-2502-16533,DBLP:journals/corr/abs-2503-15650} for details.

%,DBLP:journals/corr/abs-2503-15650

\paragraph{Statistical learning problems on graphs}
A variety of prediction problems on graphs is relevant in practice. We consider only the most prominent learning problems on graphs, which have also been theoretically studied in the literature.
Statistical learning on graphs can broadly be distinguished into two settings. The first setting comprises problems where one has information about multiple graphs (training samples), and a predictor is learnt to make predictions on new graphs; we call this \emph{a graph-level task}, which includes examples such as the classification of different proteins or molecular structures. The second setting involves a single large graph, and the goal is to make \emph{predictions about the nodes or edges of the graph}. Examples include predicting the political affiliations of individuals based on their interactions in a social network (node classification) or recommending specific products to buyers by estimating the probability of purchase (edge/link prediction).

A prototypical graph-level supervised learning problem is posed as follows. One has access to a dataset $\{(G_j, y_j):j=1,\ldots,m\}$, where each $G_j$ is a graph (for example, a protein structure) and $y_j$ is a class label or a real-valued measurement associated with $G_j$. The statistical framework assumes that $\{(G_j, y_j)\}_{j=1}^m$ are independent and identically distributed (i.i.d.) samples from an unknown joint distribution $\mathcal{D}$. Given a specified loss function, e.g., squared loss, the goal is to learn a predictor $G\mapsto f(G)$ that minimises the risk or generalisation error
$\mathcal{R}(f) = \mathbb{E}_{(G, y) \sim \mathcal{D}} \big[ \text{loss}(y,f(G))\big]$.
A generic approach to learning an optimal $f$ is to assume a hypothesis class of predictors and minimise the empirical risk: $\widehat{\mathcal{R}}(f) = \frac1m \sum_{j=1}^m \text{loss}(y_j,f(G_j))$ \cite[see][]{Bach-book-LFTP}.
In the context of GNN, the following parameterisation is assumed: a non-linear GNN learns the graph representation $G \mapsto H_G$ (Figure \ref{fig:GNN}), and a simple linear or logistic model is fitted to the representation, such as $f(G)=H_G\beta$. The predictor $G\mapsto f_\theta(G)$ is parameterized by $\theta=\left(\beta,\{W_\ell,v_\ell\}_{\ell=0,\ldots,L-1}\right)$, comprising the weights in \eqref{eqn:GCN}–\eqref{eqn:GAT} and a linear probe $\beta$.

The second setting pertains to a single large graph with $n$ nodes. We discuss only \emph{node-level prediction}, where one assumes that every node-$i$, $i=1,\ldots,n$, has an associated label or value $y_i$. Labels are only observed for $m \ll n$ nodes, and the goal is to predict the unseen labels. This setting is known as \emph{transductive inference} and is also studied in statistics as random graph models with partially labelled communities \cite{DBLP:journals/jmlr/CaiLR20}.
The GNN approach parameterizes the predictor node-$i \mapsto f(i)$ as an $L$-layer GNN that models the node representation node-$i \mapsto H^{(L)}_i$, and a simple linear or logistic probe is learnt on $H^{(L)}_i$, such as $f(i) =  H^{(L)}_i\beta$. All parameters $\theta=\left(\beta,\{W_\ell,v_\ell\}_{\ell}\right)$ are learnt by minimising a suitable training loss defined on the nodes with observed labels.

\section{Learning theoretic bounds on generalisation error}
\label{sec:gen_bounds}

We introduce some concepts of learning theory and refer to \cite{Bach-book-LFTP} for details.
Consider the graph-level prediction setting.
A hypothesis class $\mathcal{F}$ of predictors is assumed; here, it refers to $\mathcal{F}= \left\{f_\theta: \theta= \left(\beta,\{W_\ell,v_\ell\}_\ell\right)\right\}$, that is, predictors corresponding to a specific GNN architecture (fixed depth $L$, layer widths), but with all possible trainable parameters.
Given a dataset $\{(G_j, y_j)\}_{j=1}^m\sim_{i.i.d} \mathcal{D}$, it is assumed that we can obtain an \emph{empirical risk minimiser} $\widehat{f} = \text{arg\,}\min_{f\in \mathcal{F}}~\widehat{\mathcal{R}}(f)$. The goal is to derive an upper bound on the generalisation error $\mathcal{R}(\widehat{f})$ of the model $\widehat{f}$, which is generally obtained in terms of the complexity of the class $\mathcal{F}$, typically characterised by its Vapnik-Chervonenkis (VC) dimension, Rademacher complexity, etc. \cite{Bach-book-LFTP}.
For binary classification of graphs, the richness of $\mathcal{F}$ can be expressed in terms of its empirical Rademacher complexity 
$\displaystyle \widehat{\mathfrak{R}}_m(\mathcal{F}) = \mathbb{E} \left[\sup_{f \in \mathcal{F}} \left| \frac{1}{m} \sum_{i=1}^m \varepsilon_i f(G_i) \right|\right]$; the ability of $\mathcal{F}$ to fit random labels $\varepsilon_1, \dots, \varepsilon_m \sim_{i.i.d} \text{Rademacher}(\{\pm1\})$ on the training data $G_1,\ldots,G_m$.
Theorem \ref{thm:gnn_margin_bound} provides a generalisation error bound for GNNs for graph-level prediction.

\begin{theorem}[Graph-level generalisation error bound for GNNs \cite{10.5555/3524938.3525258}]
Let $\mathcal{F}$ denote the hypothesis class  for a specific GNN architecture, and let $\widehat{f}\in\mathcal{F}$ denote the empirical risk minimiser, that is, the GNN that minimises the training error $\widehat{\mathcal{R}}(f)$ given $m$ training samples $(G_j, y_j) \sim_{i.i.d.} \mathcal{D}$. With probability at least $1-\delta$, with respect to the training samples,
\begin{equation*}
\mathcal{R}(\widehat{f}\,) \leq
\widehat{\mathcal{R}}(\widehat{f}\,) +  O\left(\widehat{\mathfrak{R}}_m(\mathcal{F}) + \sqrt{\frac{\log(1/\delta)}{m}}\right) \leq
\inf_{f\in\mathcal{F}} \mathcal{R}(f) + O\left( \widehat{\mathfrak{R}}_m(\mathcal{F}) + \sqrt{\frac{\log(1/\delta)}{m}}\right).
\end{equation*}
\label{thm:gnn_margin_bound}
\end{theorem}

Theorem~\ref{thm:gnn_margin_bound} decomposes the error bound into two factors: (i) the \emph{inductive bias} $\inf_{f\in\mathcal{F}} \mathcal{R}(f)$, the minimum error incurred using the specific GNN architecture, and (ii) the \emph{estimation error} of learning the optimal predictor using finite samples, which is bounded by the Rademacher complexity.
For a $L$-layer GNN and $d$-dimensional embedding in each layer, the empirical Rademacher complexity $\widehat{\mathfrak{R}}_m(\mathcal{F}) = \mathcal{O}\left( \frac{d \cdot L \cdot b}{\sqrt{m}} \right)$, where $b$ is the maximum node degree of any graph $G_1,\ldots,G_m$ \cite{10.5555/3524938.3525258}. Thus, $m\gg dL$ samples are needed to achieve a low estimation error with a specific GNN.
Alternative error bounds using VC-dimension, margin bounds, and PAC-Bayesian analysis are also known \cite[see][]{DBLP:journals/corr/abs-2503-15650}.
Theorem \ref{thm:gnn_margin_bound} illustrates the bias-variance trade-off: a complex GNN architecture reduces the inductive bias, but the complexity of $\mathcal{F}$ and the estimation error worsen, indicating over-fitting on the training data.
To characterise the inductive bias and the complexity of $\mathcal{F}$, we need to turn towards the graph isomorphism problem.

% removed DBLP:journals/nn/ScarselliTH18, DBLP:journals/nn/DInvernoBS25, li2025towards, maskey2026graph

\paragraph{The role of graph isomorphism in graph-level GNN generalisation}

\begin{figure}[t]
\centering
\includegraphics[width=0.9\linewidth]{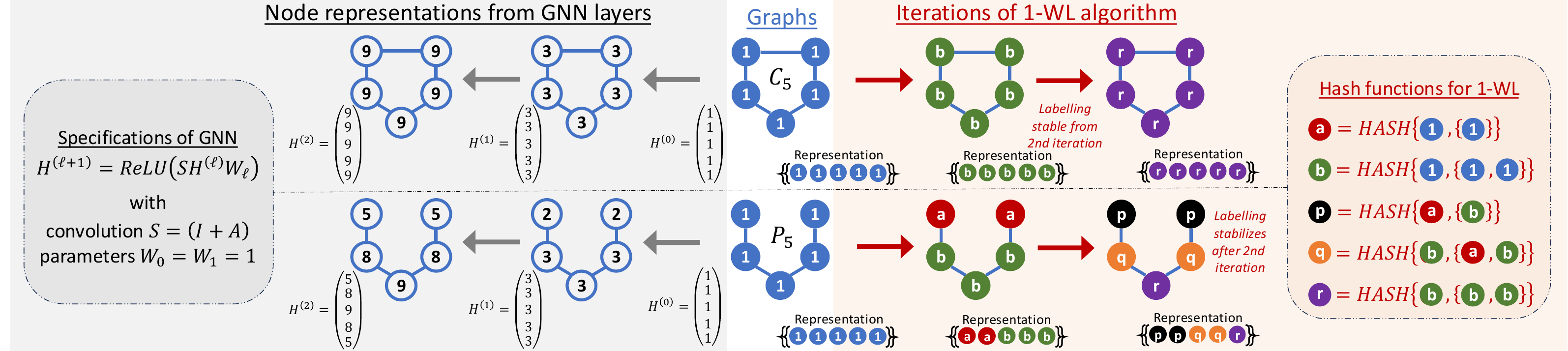}%
\caption{Illustration of how the GNN layers (left) operate on the 5-cycle graph, $C_5$, and the path graph with 5 nodes, $P_5$. For the GNN, we assume that node features are 1 for every node. 
The illustration on right compares the GNN updates with iterations of the Weisfeiler-Leman algorithm (1-WL) for checking graph isomorphism. Further GNN specifications and the Hash functions in 1-WL are noted. 
}
\label{fig:GNN-WL}
\end{figure}

The inductive bias in Theorem \ref{thm:gnn_margin_bound} characterises the expressive power of the GNN architecture. For graph-level prediction, it is equivalent to asking: \emph{Which graphs can be distinguished by a GNN under the optimal choice of parameters?} 
This relates to the graph isomorphism problem: given two graphs $G_1,G_2$, each with $n$ nodes, test whether $G_1$ and $G_2$ are isomorphic or not. Graph theory provides results on graph isomorphism, which have gained recent attention due to their relation to the expressive power of GNNs. Notably, the Weisfeiler-Leman (WL) graph isomorphism test \cite{Weisfeiler:1968ve} is closely related to GNNs.
The idea behind the WL test is to generate a multi-set representation of each graph by iteratively aggregating over the neighbourhoods and subsequently testing graph isomorphism by comparing the representations. This iterative algorithm, called 1-WL or colour refinement, proceeds as follows. Let $c^{(t)}_i$ denote the colour of node-$i$ in iteration-$t$. Initially, all nodes are identically coloured, $c^{(0)}_1=\cdots=c^{(0)}_n= c$. The colour of node-$i$ is updated as
$c^{(t+1)}_i = \text{HASH} \left( c^{(t)}_i, \{\!\{ c^{(t)}_j : (i,j) \text{ are adjacent} \}\!\} \right)$,
where $\{\!\{\dots\}\!\}$ denotes a multiset, and HASH function uniquely maps each multiset to a new colour. %\todo{add multiset notation in the fig}
The iterations repeat until the colouring stabilises.
As shown in Figure \ref{fig:GNN-WL}, the 1-WL iterations resemble the update of node representations $H^{(\ell)}$ across the GNN layers.
This equivalence has been systematically studied by several works \cite{10.1609/aaai.v33i01.33014602,DBLP:conf/iclr/GeertsR22,morris2023wl}, thereby establishing the limitations of several GNN architectures.
For example, the GCN architecture \cite{DBLP:conf/iclr/KipfW17} uses only the 1-hop neighbourhood, it is limited by the power of the 1-WL algorithm and, hence, cannot distinguish between $C_{10}$ (10-cycle) and $C_5\cup C_5$ (two disjoint 5-cycles). This has led to the proposal of new, more expressive GNNs that aggregate over larger neighbourhoods \cite{10.1609/aaai.v33i01.33014602}. However, \cite{DBLP:conf/iclr/GeertsR22} shows that the distinguishing capabilities of any MP-GNN are limited by the $k$-WL algorithm. 
The equivalence to WL not only characterises the expressive power (or inductive bias) of GNNs, but also helps to study the complexity of the hypothesis class $\mathcal{F}$ in Theorem \ref{thm:gnn_margin_bound}.
This is formalised in \citep{morris2023wl}, which relates the VC-dimension of  $\mathcal{F}$ to the WL expressivity.

% removeed GIN paper ref: DBLP:conf/iclr/XuHLJ19,

\paragraph{Generalisation error bounds in node-level prediction}

The study of generalisation error requires a different perspective when one focuses on node-level prediction, as graph isomorphism-based analysis is not needed when problems are defined on a single graph.
We discuss this using the binary node classification problem.
Given a graph $G= (X,A)$ on $n$ nodes, it is assumed that each node-$i$ has a true label $y_i \in \{\pm1\}$, drawn from a statistical law. Only the labels of $m< n$ nodes are known (assume $y_1,\ldots,y_m$ known), and the goal is to predict the labels of the remaining $n-m$ nodes. A GNN learns a map node-$i\mapsto f_\theta(i) \in \mathbb{R}$ with parameters $\theta = (\beta,\{W_\ell,v_\ell\}_\ell)$. The hypothesis class corresponding to the GNN architecture is defined as $\mathcal{F} = \left\{ f_\theta \in \mathbb{R}^n : \theta = (\beta,\{W_\ell,v_\ell\}_\ell)\right\}$.
For some specified loss function, such as 0-1 error, the empirical risk and the generalisation error are defined as $\widehat{\mathcal{R}}(f) = \displaystyle\frac1m \sum_{i=1}^m \text{loss}(y_i,f(i))$ and ${\mathcal{R}}(f) = \displaystyle\frac1n \sum_{i=1}^n \text{loss}(y_i,f(i))$, respectively, and one aims to derive guarantees akin to Theorem \ref{thm:gnn_margin_bound}.
In this context, the complexity of the class $\mathcal{F}$ can be characterised in terms of the \emph{transductive Rademacher complexity} \citep{el2009transductive}, defined as
$\displaystyle \widehat{\mathfrak{R}}_{m,n}(\mathcal{F}) = \left( \frac{1}{m} + \frac{1}{n-m} \right) \mathbb{E}_{\varepsilon_1,\ldots,\varepsilon_n} \left[ \sup_{f \in \mathcal{F}} \sum_{i=1}^n \varepsilon_i f(i) \right]$,
where $\varepsilon_i \in\{-1,0,+1\}$ are i.i.d. variables that are Rademacher with probability $p\in(0,1)$ and 0 with probability $1-p$, indicating observed and unobserved labels. The following theorem from \cite{DBLP:conf/nips/EsserVG21} shows that the generalisation error in node level prediction is influenced by the choice of convolution $S$, as well as its interaction with the node features through $SX$. 
%An alternative bound for a 1-layer GCN in terms of the spectrum of $S$ is given in \cite{lv2021generalization}, 
\cite{DBLP:conf/nips/EsserVG21} further proves that using VC-dimension leads to trivial bounds in this context.

\begin{theorem}[Generalisation error bound for node classification using GCN \cite{DBLP:conf/nips/EsserVG21}]
Consider the node classification setting. Let $\widehat{f}\in\mathcal{F}$ be the empirical risk minimiser. Then
\begin{equation*}
\mathcal{R}(\widehat{f}\,) \leq \widehat{\mathcal{R}}(\widehat{f}\,) + \widehat{\mathfrak{R}}_{m,n}(\mathcal{F}) + O\left(\max\left\{ \frac{1}{\sqrt{m}}, \frac{1}{\sqrt{n-m}}\right\}\cdot\left(1+ \sqrt{\log(1/\delta)}\right)\right),
\end{equation*}
with probability at least $1-\delta$. 
Furthermore, for a $L$-layer GCN \eqref{eqn:GCN} with convolution operator $S$ and Lipschitz smooth activation, the transductive Rademacher complexity is bounded as
\begin{equation}
\widehat{\mathfrak{R}}_{m,n}(\mathcal{H}) \le \mathcal{O}\left( \frac{n^2}{m(n-m)} \left( \sum_{\ell=0}^{L-1} \|S\|_\infty^\ell \right)+ \|S\|_\infty^L \|SX\|_{2\to\infty} \sqrt{\ln n}  \right),
\end{equation}
where $\Vert\cdot\Vert_\infty$ in the induced $\infty$-norm and $\Vert\cdot\Vert_{2\to \infty}$ is the entry-wise $2\to\infty$-norm of a matrix.
\label{thm:gnn_transductive_bound}
\end{theorem}

\section{Deep Learning Asymptotics of GNNs}
\label{sec:asymptotics}

The above generalisation error bounds are not adapted to the training process, resulting in loose bounds that do not explain the empirical behaviour of trained GNNs. Notably, they cannot explain \emph{why many GNNs suffer from oversmoothing}, i.e., their performance degrades drastically with depth (increasing  $L$); or \emph{why certain GNNs perform well on homophilic graphs but not on heterophilic graphs}. 
To answer these questions, one needs to characterise the GNN learnt via training: a challenging problem for nonlinear neural networks. Hence, either generalisation error is studied for linear GNN \cite{10.5555/3600270.3600435,ayday2025why,doi:10.1073/pnas.2309504121}, or the GNN is simplified under different asymptotic regimes. 

Studying neural networks in the limit of infinitely-many parameters has been a useful tool in deep learning theory, as it provides tractable characterisations of the limiting model that retain many statistical properties of the finite neural networks used in practice, such as the double descent phenomenon, scaling laws, etc. The asymptotics can be achieved at an infinite depth $(L\to\infty)$, at an infinite hidden-layer width $(\min_{\ell\in\{1,\ldots,L\}} d_\ell \to\infty)$, and with infinite attention heads in attention-based models (like GAT), among others.
Depth asymptotics, $L\to\infty$, in GNNs lead to trivial representations due to oversmoothing.\footnote{This is easily verified in linear GCN. Here, the update rule \eqref{eqn:GCN} with $\sigma$ as the identity map simplifies to $H^{(L)} = S^L X W_{L-1}\cdots W_1$. For the normalised convolution operators, matrix algebraic calculations show that $S^L$ converges to a rank one matrix as $L\to\infty$, and hence, GCN cannot distinguish between nodes \cite{10.5555/3600270.3600435}.}  This has been rigorously studied from a generalisation error perspective in \cite{DBLP:conf/iclr/OonoS20,10.5555/3600270.3600435}, and approaches for mitigating oversmoothing via residual connections or multi-scale representations have been analysed \cite{DBLP:conf/nips/OonoS20,sabanayagam2023analysis}. We discuss two asymptotic regimes that give useful characterisations of the trained GNN.

\begin{figure}[t]
    \centering
    \begin{minipage}{0.58\textwidth}
    \includegraphics[width=0.32\linewidth]{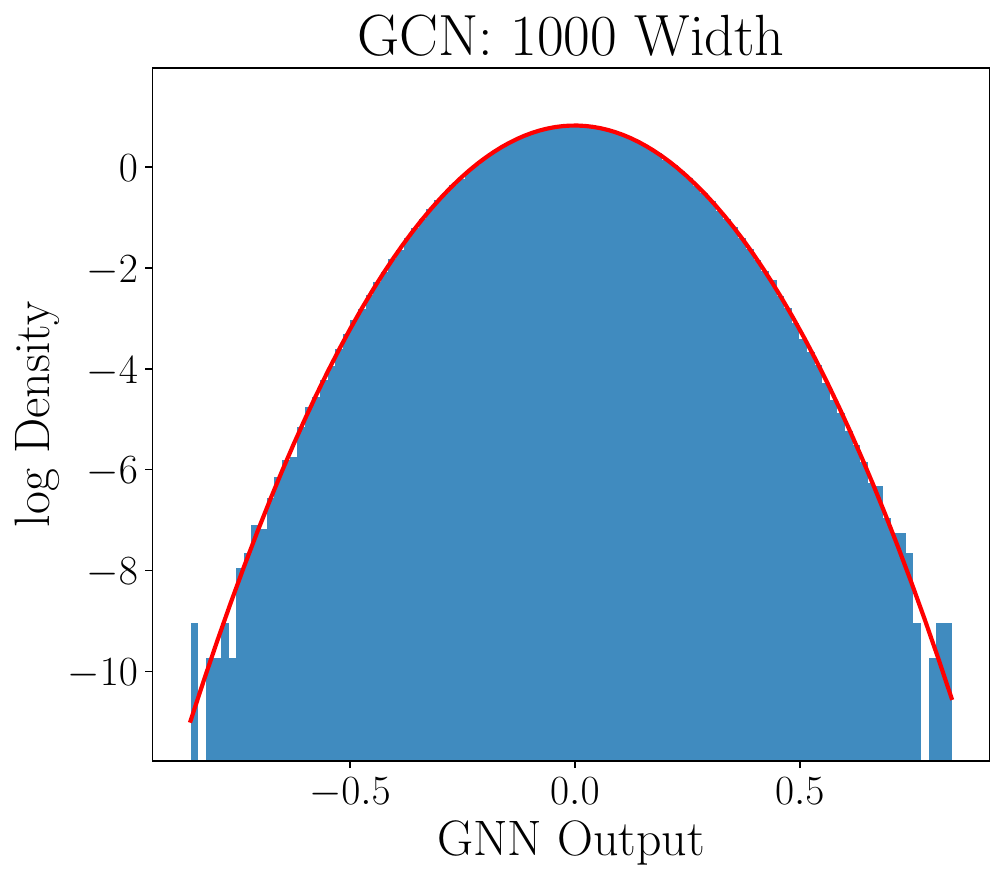}   
    \includegraphics[width=0.32\linewidth]{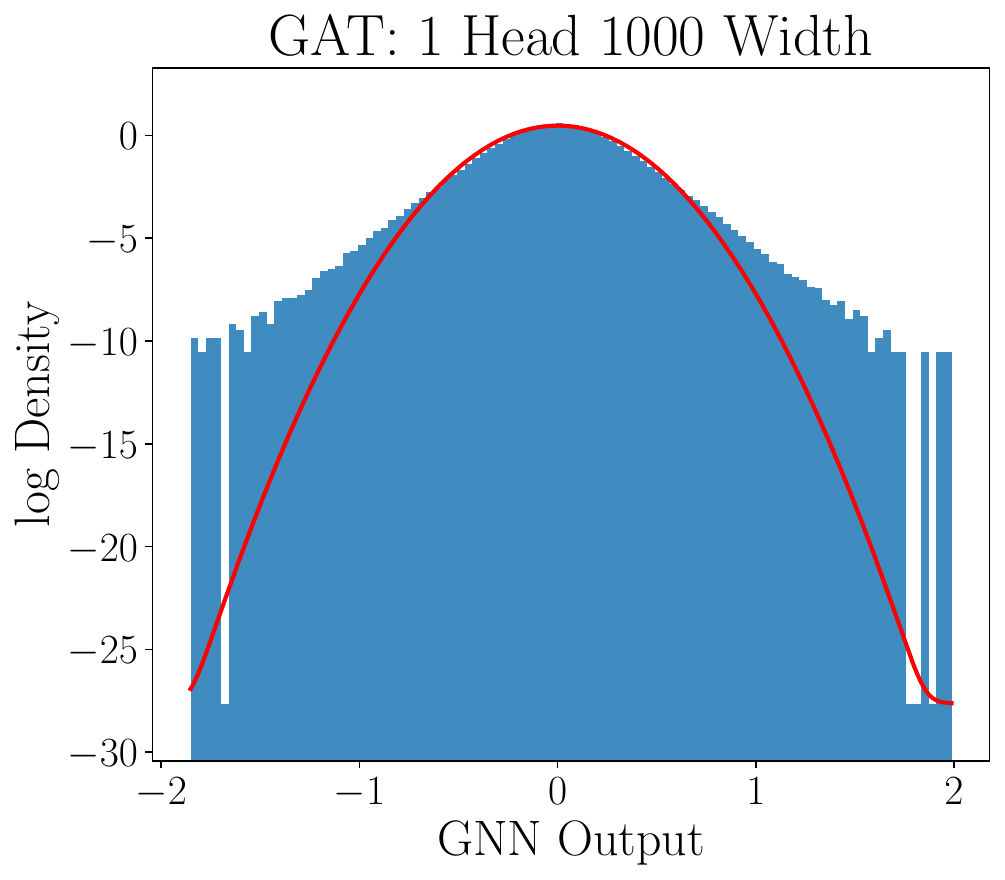}
    \includegraphics[width=0.32\linewidth]{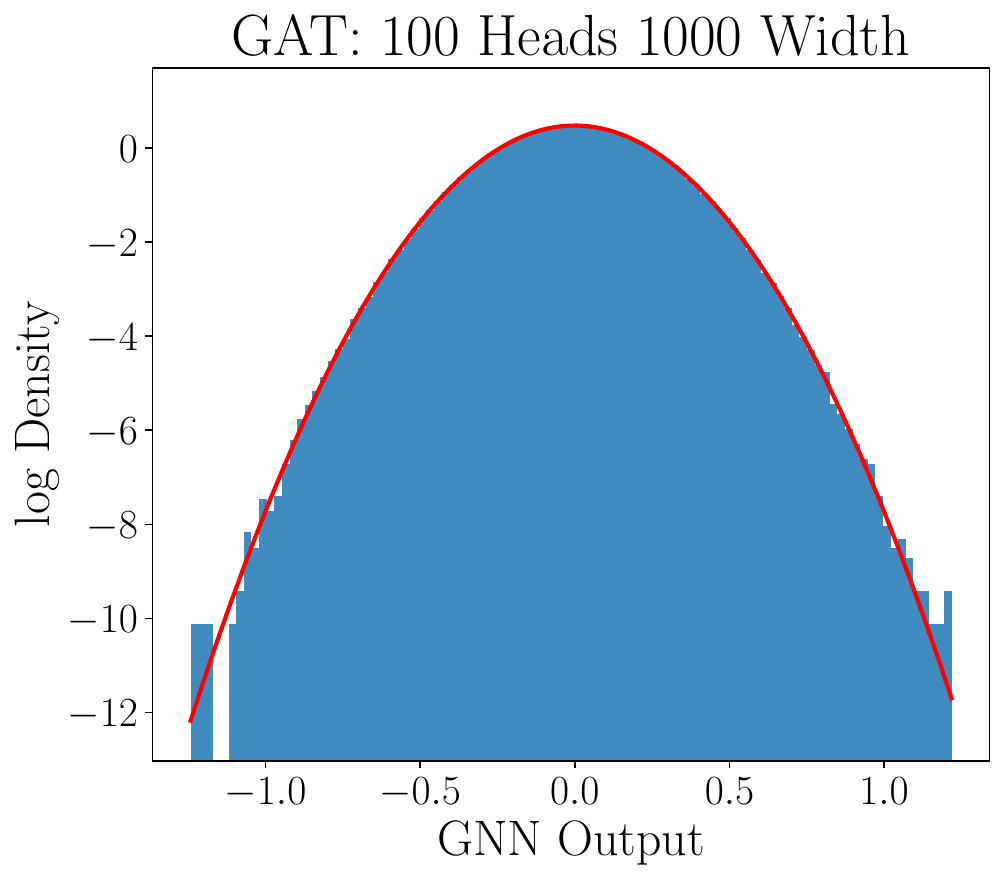}
    \caption{Histograms of $\{f_{\theta_0}(i)\}_{i=1,\ldots,n}$ for randomly initialised $\theta_0$ in 1-layer GCN and GAT for Erd\H{o}s-R{\'e}nyi graphs with $n=100$ \cite{DBLP:journals/corr/abs-2603-17569}. {\bf(left)} For GCN with width $d_i=1000$, histogram of $f_{\theta_0}(i)$ is close to a Gaussian distribution (red line fitted with mean and variance). {\bf(middle)} For GAT in \eqref{eqn:GCN}--\eqref{eqn:GAT}, the histogram is not Gaussian. {\bf(right)} In practice, each layer of GAT learns multiple parallel attention heads $S^{(\ell)}$. GAT with 1000 width and 100 heads shows a GP behaviour, indicating that GP-limit holds when width and number of attention heads are simultaneously infinite.}
    \label{fig:gcn_gat_comparison}
    \end{minipage}     
    \hfill
    \begin{minipage}{0.36\linewidth}
    \centering
    \includegraphics[width=0.98\linewidth]{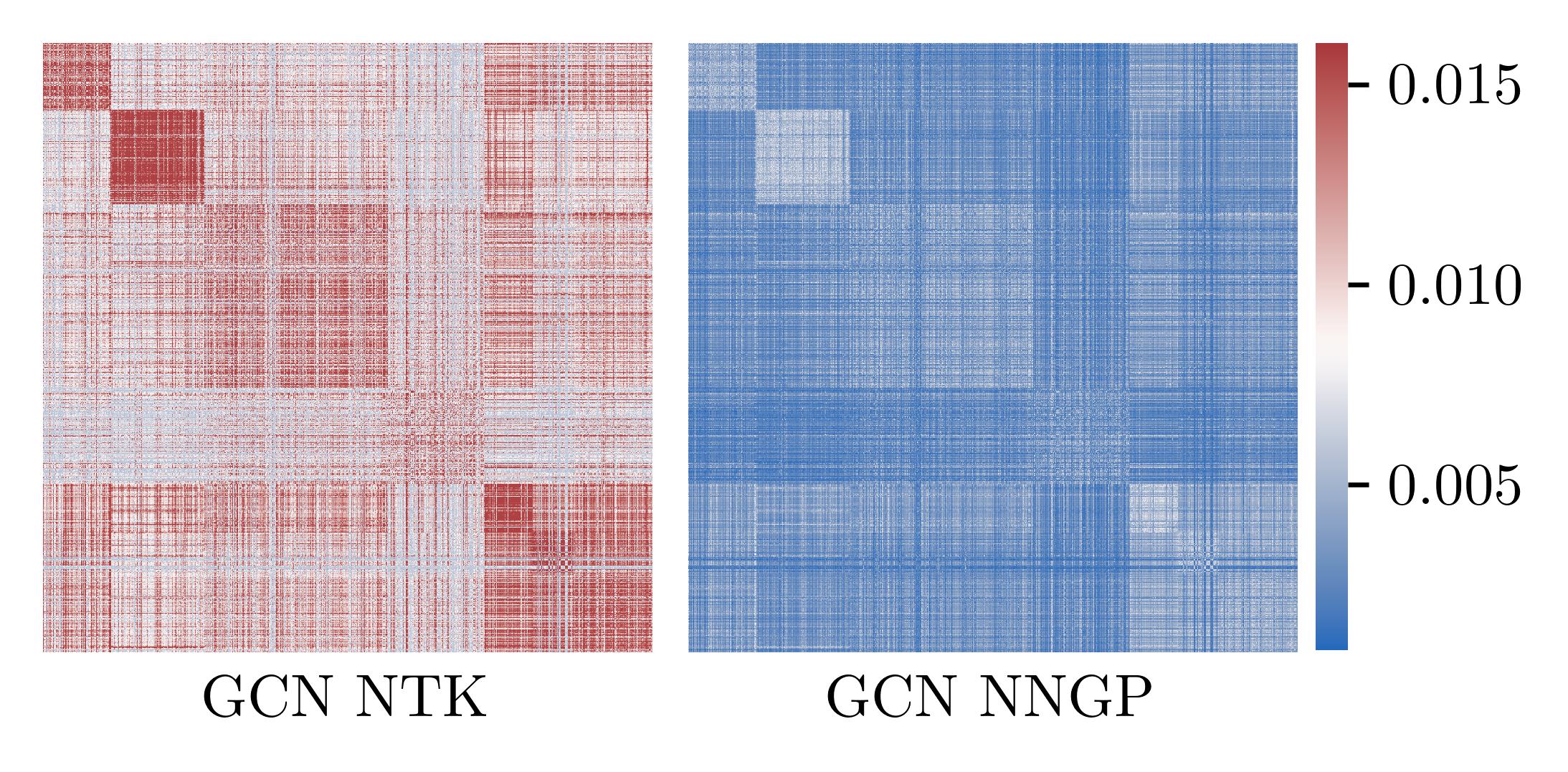}
    \caption{Comparison of the kernels for Graph NTK {\bf(left)} and Graph NNGP {\bf(right)} derived for a 2-layer GCN on CORA dataset $(n=2708)$. Both kernels show a block structure, corresponding to the 7 communities in CORA graph, but the blocks are more prominent in the Graph NTK. The plots are adapted from \cite{sabanayagam2023analysis}.}
    \label{fig:gcn_gp_ntk_comparison}
    \end{minipage}
\end{figure}

\paragraph{Asymptotics of wide GNNs: Gaussian Process and Neural Tangent Kernel}

Consider node-level prediction with the GNN map node-$i\mapsto f_\theta(i)=H^{(L)}_i\beta$, where $\theta = (\beta, \{W_\ell,v_\ell\}_\ell)$. In practice, the parameters are randomly initialised $\theta=\theta_0$ and are updated via gradient descent or its variants to minimise the training error.
The aim of width asymptotics of GNN, or broadly neural networks, is to approximate the GNN $f_\theta$ by a model $\tilde{f}_\theta$ that is nonlinear with respect to the input but linear in trainable parameters; that is, $f_\theta(\cdot) \approx \tilde{f}_\theta(\cdot) = \theta^\top \phi(\cdot)$ for some non-trainable nonlinear map $\phi(\cdot)$. 

The first approximation stems from the observation that if all parameters in  $\theta = \theta_0$ are initialised as independent centred Gaussians with appropriately scaled variances, then by the central limit theorem, the map node-$i\mapsto f_{\theta_0}(i)$ converges to a Gaussian process (GP) on the nodes as $\min_{\ell\in\{1,\ldots,L\}} d_\ell \to\infty$ \cite{DBLP:conf/iclr/LeeBNSPS18} (see Figure \ref{fig:gcn_gat_comparison}). In the context of GNNs, this is referred to as a \emph{Graph NNGP}, and the corresponding covariance kernel for GCN \eqref{eqn:GCN} has been derived  in \cite{niu2023graph} (see Theorem \ref{thm:gnngp_gntk}), while \cite{DBLP:journals/corr/abs-2603-17569} derives the kernel for GAT \eqref{eqn:GCN}--\eqref{eqn:GAT} and graph transformers.
From a Bayesian perspective, the limiting GP $\{f_{\theta_0}(i)\}_{i=1,\ldots,n}$ is viewed as a prior, and the posterior prediction on the unlabelled nodes can be obtained by solving a GP regression or classification given the labels $\{y_i\}_{i=1,\ldots,m\ll n}$. One may view the NNGP as a \emph{random feature} model $\tilde{f}(i)=H_i^{(L)}\beta$, where $H_i^{(L)}$ is obtained from the random initialisation and $\beta$ is learnt; and, in the infinite-width limit, the  kernel matrix is $K= \mathbb{E}_{\theta_0}\left[H^{(L)}(H^{(L)})^\top\right] \in \mathbb{R}^{n\times n}$.
Note that the Graph NNGP is an over-simplification of GNN since only the last layer $\beta$ is learnt. 

A better approximation can be derived via Taylor expansion at the initialisation $\theta_0$ as
$f_\theta(\cdot) \approx f_{\theta_0}(\cdot) + \nabla_\theta f_{\theta_0}(\cdot)^\top (\theta - \theta_0) + \frac{1}{2}(\theta - \theta_0)^\top \nabla^2_\theta f_{\theta_0}(\cdot) (\theta - \theta_0)$.
It has been shown that \cite{liu2020linearity}, under suitable random initialisation $\theta_0$, and in the infinite-width limit, the second-order term vanishes. Hence, $f_\theta(\cdot)$ can be approximated by a kernel machine, with the kernel given by $K_{ij} = \mathbb{E}_{\theta_0}\left[(\nabla_\theta f_{\theta_0}(i))^\top \nabla_\theta f_{\theta_0}(j)\right]$, accounting for the infinite-width limit of the gradient at random initialisation.
This is called the \emph{neural tangent kernel} (NTK) \cite{jacot2018neural}, which has been derived for node and graph-level GCN in \cite{sabanayagam2023analysis,du2019gntk}.   
Theorem \ref{thm:gnngp_gntk} and Figure \ref{fig:gcn_gp_ntk_comparison} show the difference between Graph NNGP and Graph NTK.

\begin{theorem}[Node-level kernels of Graph NNGP and Graph NTK for GCN \cite{niu2023graph,sabanayagam2023analysis}]
Consider the $L$ layer GCN in \eqref{eqn:GCN} with node-level output $f_\theta(i) = H^{(L)}_i\beta$, and assume that the activation $\sigma$ is differentiable. For $\ell=0,\ldots,L$, define the matrices
$\Sigma_0 = SXX^\top S^\top$ and $\Sigma_{\ell+1} = S E_\ell S^\top$, where
$\displaystyle E_\ell = \mathbb{E}_{z \sim \mathcal{N}(0, \Sigma_\ell)}\big[\sigma(z)\sigma(z)^\top\big]$ and 
$\displaystyle \dot{E}_\ell = \mathbb{E}_{z \sim \mathcal{N}(0, \Sigma_\ell)} \big[\dot\sigma(z)\dot\sigma(z)^\top\big]$.
The node-level Graph NNGP kernel is given by $K_{nngp} = E_{L} \in \mathbb{R}^{n \times n}$ and the Graph NTK is %given by
\[
K_{ntk} = \sum_{k=0}^{L-1} \underbrace{S \Big( \ldots S \Big( S}_{L-1-k \text{ terms}} \left( {\Sigma}_k \odot \dot{E}_k \right) S^\top \odot  \dot{E}_{k+1} \Big) S^\top \odot \ldots \Big) S^\top \odot \dot{E}_{L-1} \in \mathbb{R}^{n\times n},
\]
where $\odot$ denotes Hadamard product. The NNGP and NTK limits hold assuming that the parameters are independently initialised as $(W_\ell)_{ij}\sim\mathcal{N}(0,\frac{1}{d_\ell})$ and $d_\ell\to\infty$ for $\ell=1,\ldots,L$.
\label{thm:gnngp_gntk}
\end{theorem}

The NNGP and NTK limits allow for approximating the complex GNNs with kernel machines, where the optimal parameters can be characterised, specifically in squared regression or margin-based classification. The approximate models are competitive with GNNs in practice \cite{du2019gntk,niu2023graph}, and they also provide kernel-based generalisation error bounds for GNNs \cite{du2019gntk}.
Notably, the NNGP and NTK limits retain the structure of the graph convolutions $S$, thereby making it possible to analyse the influence of the graph on GNN performance. For example, NTK shows why row-normalised convolution $S_{row}$ and skip connections improve the performance of GCNs \cite{sabanayagam2023analysis}, and NNGP-based structural comparison reveals why GATs can avoid oversmoothing at depth \cite{DBLP:journals/corr/abs-2603-17569}.
On the practical front, Graph NTK helps to design deterministic robustness certificates against data poisoning, which is the most difficult adversarial setting for GNNs \cite{sabanayagam2025exact}.

\paragraph{Asymptotics of the graph size: Graphon Neural Network}

GNNs can also be studied under the asymptotics of the graph size. The convergence of a sequence of undirected graphs $A \in [0,1]^{n\times n}$ with an increasing number of nodes $n$ is well studied. The convergence is defined in terms of cut-distance or subgraph counts, and the limiting object is a symmetric measurable function $\mathcal{A}:[0,1]^2 \to [0,1]$, known as a \emph{graphon} \cite{DBLP:books/daglib/0031021}.
Hence, for a convergent sequence of growing graphs, the associated GNNs also converge to a measurable operator. This limiting object has been independently introduced as continuous GNNs \cite{DBLP:conf/nips/KerivenBV20} and graphon neural networks \cite{DBLP:conf/nips/RuizCR20}.
The necessity of studying the graphon limit of GNNs must be clarified: while the graph remains fixed for node-level predictions, the limit allows one to examine the stability of GNN predictions \cite{DBLP:conf/nips/KerivenBV20} and the transferability of trained GNNs in graph-level task \cite{DBLP:conf/nips/RuizCR20}.

The following exposition of Graphon Neural Network builds on \eqref{eqn:GCN} and is a simplification of \cite{DBLP:conf/nips/KerivenBV20,DBLP:conf/nips/RuizCR20}.
In the limit of $n\to\infty$, the vertex set is mapped to the interval $[0,1]$, the graph adjacency converges to the graphon $\mathcal{A}:[0,1]^2\to[0,1]$, and the node representation at layer-$\ell$ is a map $h^{(\ell)}:[0,1]\to \mathbb{R}^{d_\ell}$, viewed as $d_\ell$ measurable functions $h^{(\ell)}(\cdot) = [h^{(\ell,1)}(\cdot) \cdots h^{(\ell,d_\ell)}(\cdot)]$. The GCN \eqref{eqn:GCN} extends to graphon neural network as
\[
h^{(\ell+1)} = \sigma\left(\left[\mathcal{S}(h^{(\ell,1)}) \cdots \mathcal{S}(h^{(\ell,d_\ell)})\right]W_\ell\right), ~\mathcal{S}(h)(x) = \int_0^1 \frac{\mathcal{A}(x,x')}{\sqrt{\text{deg}(x)}\sqrt{\text{deg}(x')}}h(x') \textrm{d}x'.
\]
The above convolution operator $\mathcal{S}$ on measurable functions extends the symmetric normalised convolution $S_{sym}$, and $\text{deg}(x) = \int_0^1 \mathcal{A}(x,x')\textrm{d}x'$ is the degree function. 
We do not make assumptions about node features, which could be viewed as a measurable function $h^{(0)}:[0,1]\to\mathbb{R}^{d_0}$. Hence, the graphon neural network is an operator mapping $h^{(0)}\mapsto h^{(L)}$. One may also obtain the graph level representation via mean pooling as $h_G = \int_0^1 h^{(L)}(x)\textrm{d}x$.
\cite{DBLP:conf/nips/KerivenBV20,DBLP:conf/nips/RuizCR20} provide rates for the convergence of node- and graph-level representation from  discrete $n$-node GNNs to continuous graphon neural networks. For finite width GNNs and dense graphs, the convergence rates are $O(n^{-1/2})$ \cite{DBLP:conf/nips/RuizCR20}.  \cite{krishnagopal2023graph} studies the asymptotics under both width and graph size, deriving a \emph{Graphon NTK}.

\section{Study of GNNs under Random Graph Models}
\label{sec:graph_model}

The study of random graphs has been a central theme at the intersection of statistics and network analysis. Specifically, analysis under specific random graph models has provided deep insights into the statistical and computational hardness of graph problems, such as community detection, and has also led to important tools and techniques, such as message-passing algorithms and the non-backtracking operator \cite{DBLP:journals/jmlr/Abbe17,DBLP:journals/jmlr/CaiLR20,DBLP:conf/iclr/ChenLB19}. Interestingly, the statistical analysis of GNNs under random graphs is relatively unexplored, with only a few works providing concrete statistical guarantees in this setting. 

The \emph{Contextual Stochastic Block Model} (CSBM) \cite{DBLP:conf/icml/BaranwalFJ21,DBLP:journals/jmlr/FountoulakisLYBJ23} is well suited for studying GNNs. 
In a CSBM$(n,p,q,\text{\boldmath$\mu$})$, the $n$ nodes are split into two latent communities with labels $y_1,\ldots,y_n \sim_{i.i.d} \text{Rademacher}(\{\pm1\})$. The node features $X\in\mathbb{R}^{n\times d_0}$ are sampled from a Gaussian mixture model with $X_i \sim \mathcal{N}(y_i\text{\boldmath$\mu$}, I_{d_0\times d_0})$, where $\text{\boldmath$\mu$} \in \mathbb{R}^{d_0}$ is specified. The graph $A\in\{0,1\}^{n\times n}$ has independent undirected edges $A_{ij}\sim\text{Bernoulli}(p)$ for $y_i=y_j$, and $A_{ij}\sim\text{Bernoulli}(q)$ for $y_i \neq y_j$, where $0< q<p<1$ are specified.
We state the following result, adapted from \cite{DBLP:conf/icml/BaranwalFJ21,DBLP:journals/jmlr/FountoulakisLYBJ23}, which considers a simplification of the node classification problem. The authors study 1-layer GCN \eqref{eqn:GCN} and 1-layer GAT \eqref{eqn:GCN}--\eqref{eqn:GAT} with fixed parameters $W_0, v_0, \beta$ and verify if the map node-$i\mapsto f(i)= H_i^{(1)}\beta$ has the same sign as $y_i$. We use the standard asymptotic notations $O(\cdot),\Omega(\cdot),\omega(\cdot)$.

\begin{theorem}[Exact recovery of latent classes in CSBM using GCN or GAT \cite{DBLP:conf/icml/BaranwalFJ21,DBLP:journals/jmlr/FountoulakisLYBJ23}]
Consider a random graph $(X,A) \sim \textup{CSBM}(n, p, q, \text{\boldmath$\mu$})$. Assume that $p,q =\omega\left(\frac{(\log n)^2}{n}\right)$ and $\frac{p-q}{p+q}=\Omega(1)$.

\noindent\underline{Analysis of GCN:} If $\Vert\text{\boldmath$\mu$}\Vert = \omega\Big(\frac{\log n}{\sqrt{n(p+q)}}\Big)$, there is a 1-layer GCN with sigmoid activation and specific $\theta=(W_0,\beta)$ so that $\textup{sign}(f_\theta(\cdot))$ recovers labels $y_1,\ldots,y_n$ with probability $1-o(1)$.

\noindent\underline{Analysis of GAT:} If $\Vert\text{\boldmath$\mu$}\Vert = \omega\left(\sqrt{\log n}\right)$, there exist parameters $\theta=(W_0,v_0,\beta)$ for a 1-layer GAT with leaky ReLU activation in \eqref{eqn:GAT} such that, with probability $1-o(1)$, $\textup{sign}(f_\theta(\cdot))$ exactly recovers all labels. However, the model mis-classifies $\Omega(n)$ nodes if $\Vert\text{\boldmath$\mu$}\Vert = o\left(\sqrt{\frac{\log n}{n(p+q)}}\right)$.

\noindent\underline{Performance ignoring the graph $A$:} As a baseline, it can be shown that if only the node features $H^{(0)}=X$ are used, then if $\Vert\text{\boldmath$\mu$}\Vert = \omega\left(\sqrt{\log n}\right)$, then there exists a linear classifier $f_\beta(i)=\textup{sign}(H_i^{(0)}\beta)$ that exactly recovers all labels with probability $1-o(1)$. However, if 
$\Vert\text{\boldmath$\mu$}\Vert = O\left(1\right)$, then $\big\{H_i^{(0)}\big\}_{i=1,\ldots,n}$ are not separable by linear classifiers with probability $\Omega(1)$.
\label{thm:CSBM}
\end{theorem}

Theorem \ref{thm:CSBM} shows that for semi-sparse graphs, $p,q = \omega\big( \frac{(\log n)^2}{n}\big)$, a 1-layer GCN can exactly recover communities even when the node features $X$ cannot be linearly separated. The statement for GAT is nuanced, as weak separation of features (small {\boldmath$\mu$}) leads to poor learning of the attention $S^{(0)}$. However, Theorem \ref{thm:CSBM} assumes fixed parameters and does not account for training the GNN. 
The only works that study the statistical performance of GNNs with learnt parameters are \cite{10.5555/3600270.3600435,ayday2025why,doi:10.1073/pnas.2309504121}, where the analysis is restricted to GCNs with linear activation. In particular, \cite{doi:10.1073/pnas.2309504121} precisely derives the generalisation error of linear GCN for node-level classification under a CSBM, thereby showing that GCN exhibits a double descent phenomenon, which is modulated by the graph signal-to-noise ratio $\frac{n(p-q)}{\sqrt{n(p+q)}}$.
Observe that node features $X$ and adjacency $A$ are  aligned in CSBM; hence, in Theorem \ref{thm:CSBM}, GCN can improve upon the linear classification of $X$. This improvement does not universally hold in practice. To study the effect of misalignment, \cite{ayday2025why} derives the generalisation error of linear GCN under a more general model, where the weighted graph may not completely align with the node features. The error in \cite{ayday2025why} identifies failure modes of GNN architectures, stemming from misalignment or from heterophilic graph structures. \cite{10.5555/3600270.3600435} introduces a contextual variant of the graphon model and derives generalisation error bounds for linear GCN.

\section{Discussion and Open Problems}

In this paper, we provide an overview of the different approaches for studying statistical generalisation in GNNs for graph-level and node-level prediction problems, exemplified through Theorems \ref{thm:gnn_margin_bound}--\ref{thm:CSBM}. To summarise, the learning theoretic perspective (Section \ref{sec:gen_bounds}) provides data-agnostic generalisation error bounds for GNNs. In conjunction with expressivity results for GNNs, these bounds can be useful for graph-level prediction. However, the bounds are not tight and may not reflect the empirical performance of trained GNNs. 
Asymptotic analysis of GNNs (Section \ref{sec:asymptotics}), specifically the asymptotics in the network width, the attention heads, or the graph size, helps to simplify them to either kernel machines or operators on function space. This helps to study the statistical performance or the stability of trained GNNs, albeit remaining agnostic to the data or graph model.
The study of GNNs under specific random graph models (Section \ref{sec:graph_model}), such as CSBM, allows for a data-dependent analysis of statistical error in GNNs. This direction is emerging, and there are no statistical guarantees for trained GNNs beyond the specific case of linear GCNs.
We note some interesting research directions and open questions, focusing on graph-, node-, or edge-level prediction problems.

\smallskip\noindent\emph{Node-level prediction.} 
The statistics literature has made tremendous contributions to the community detection problem \cite{DBLP:journals/jmlr/Abbe17,DBLP:journals/jmlr/CaiLR20}, introducing CSBM and other models. While node-level prediction using GNN is closely related, the current GNN guarantees do not compare to the optimal estimators developed in statistics (for example, Theorem \ref{thm:CSBM} holds only for semi-sparse graphs). More importantly, existing works do not provide precise error rates for trained GNNs, beyond the simple case of linear GNNs. 

\begin{quote}
\emph{Research questions}: 
The central open problem is to derive the generalisation error of a trained non-linear GNN under sparse CSBM with partial labels. This would answer if GNNs are optimal for CSBM in comparison to spectral or message passing estimators \cite{DBLP:journals/jmlr/Abbe17,DBLP:journals/jmlr/CaiLR20}. One may also study whether GNNs provide benefits under general latent space models \cite{ayday2025why,10.5555/3600270.3600435}.
\end{quote}

\noindent
Studying trained GNNs requires accounting for optimisation. Recent work on neural networks has investigated this problem for learning in high dimensions \cite{DBLP:conf/colt/AbbeAM23}, but there are no similar studies on GNNs. An alternative is to use the NNGP or NTK limits of GNNs and study their error rates under CSBM. However, this is complicated since existing convergence guarantees either require fixed $n$ for width $d_\ell\to\infty$ \cite{liu2020linearity} or fixed width for $n\to\infty$ \cite{DBLP:conf/nips/RuizCR20}. An interesting problem is to study the convergence rates to a Graphon NTK (or NNGP) as $n,d_\ell\to\infty$, which can be subsequently used to derive the non-asymptotic error rates for trained GNNs.

\smallskip\noindent\emph{Graph-level prediction.} A potential path towards reconciling the theory of GNN-based graph-level prediction with statistical network analysis would be to consider hypothesis testing of random graphs. Recent work has developed testing procedures and statistical consistency under general models (inhomogeneous Erd\H{o}s-R{\'e}nyi, graphon, exponential random graph models (ERGM)), as well as their relation to other graph learning problems \cite{DBLP:conf/iclr/SabanayagamVG22,ghosh2026maximum}. Graphon-based results for GNNs \cite{DBLP:conf/nips/KerivenBV20,10.5555/3600270.3600435,DBLP:conf/nips/RuizCR20} provide a direction to study GNNs in the context of hypothesis testing of graphs or graph classification.

\begin{quote}
\emph{Research questions}: 
Can one design a consistent GNN-based two-sample test for random graphs sampled from graphons or ERGMs? If a graphon neural network computes different representations $h_G$ for two graphons, then its discretised GNN could potentially lead to a consistent test.
On a related note, studying Weisfeiler-Leman tests \cite{DBLP:journals/corr/abs-2503-15650,DBLP:conf/iclr/GeertsR22} in the context of hypothesis testing of random graphs could relate network statistics to graph isomorphism.
\end{quote}

% removed citation  \cite[see][]{DBLP:conf/nips/BokerLHV023}

\smallskip
\noindent\emph{Edge-level prediction.}
Edge or link prediction is an important problem in statistical network analysis: Given a graph $G$ with a partially observed adjacency, the goal is to predict the unobserved links. It is also closely related to the \emph{network model estimation} problem. 
Little is known about the generalisation error of link prediction using GNNs. 
While the GNN literature typically uses the node representation $H^{(L)}$ for link prediction, a notable work \cite{DBLP:conf/nips/ZhouK022} uses the graphon limit to show that such methods can fail in out-of-distribution generalisation to larger graphs and suggests learning representations of node pairs. 
This idea is related to attention learning in GAT or transformer architectures; for example, $S^{(\ell)}$ in \eqref{eqn:GAT} provides an estimate of edge probability. \cite{DBLP:journals/corr/abs-2603-17569} derives an edge-level kernel in the GP limit of GAT, which could be used for link prediction, while the analysis of GAT with specific parameters under CSBM shows when the graph attention can or cannot predict the edges \cite{DBLP:journals/jmlr/FountoulakisLYBJ23}.

\begin{quote}
\emph{Research questions}:
There is a need to derive statistical guarantees for GNN-based link prediction under general graph models, such as graphons \cite{GaoLZ2015graphon}. Such results can both establish the suitability of GNNs for network model estimation and rigorously explain the significance of attention learning in GATs, which implicitly modifies the graph structure.   
\end{quote}

\bmhead{Acknowledgements}

The work of N. Ayday and D. Ghoshdastidar is supported by the German Research Foundation (DFG) Priority Program on Theoretical Foundations of Deep Learning (SPP-2298; GH-257/2-2). N. Ayday also acknowledges support from the Konrad Zuse School of Excellence in Reliable AI.

\bibliography{sn-bibliography}% common bib file

\end{document}